\pdfoutput=1
\documentclass[prl,twocolumn,amsmath,amssymb,aps]{revtex4-1}
\usepackage{times}
\usepackage{graphicx}
\usepackage{bm}
\usepackage{natbib}
\usepackage{textcomp}
\begin{document}
\title{Spin triplet superconducting proximity effect in a ferromagnetic semiconductor}

\author{Taketomo Nakamura${}^1$}
\email{taketomo@issp.u-tokyo.ac.jp}
\author{Le Duc Anh${}^{2,3}$}
\author{Yoshiaki Hashimoto${}^1$}%
\author{Shinobu Ohya${}^{2,3,4}$}
\author{Masaaki Tanaka${}^{3,4}$}
\author{Shingo Katsumoto${}^{1,4}$}%
\affiliation{${}^1$Institute for Solid State Physics, University of Tokyo, 5-1-5 Kashiwanoha, Kashiwa, Chiba 277-8581, Japan}
\affiliation{${}^2$Department of Electrical Engineering and Information Systems,University of Tokyo, 8-3-1 Hongo, Bunkyo, Tokyo 113-8656, Japan}
\affiliation{${}^3$Institute of Engineering Innovation, Graduate School of Engineering, The University of Tokyo, 7-3-1 Hongo, Bunkyo-ku, Tokyo 113-8656, Japan}
\affiliation{${}^4$Center for Spintronics Research Network (CSRN), The University of Tokyo, 7-3-1 Hongo, Bunkyo-ku, Tokyo 113-8656, Japan}

\date{\today}
\begin{abstract}
Conventional spin-singlet superconductivity that deeply penetrates into ferromagnets is typically killed by the exchange interaction, which destroys the spin-singlet pairs. Under certain circumstances, however, superconductivity survives this interaction by adopting the pairing behavior of spin triplets. The necessary conditions for the emergence of triplet pairs are well-understood, owing to significant developments in theoretical frameworks and experiments.
The long-term challenges to inducing superconductivity in magnetic semiconductors, however, involve difficulties in observing the finite supercurrent, even though the generation of superconductivity in host materials has been well-established and extensively examined.
Here, we show the first evidence of proximity-induced superconductivity in a ferromagnetic semiconductor (In, Fe)As. The supercurrent reached a distance scale of $\sim 1~\mu$m, which is comparable to the proximity range in two-dimensional electrons at surfaces of pure InAs.
Given the long range of its proximity effects and its response to magnetic fields, we conclude that
spin-triplet pairing is dominant in proximity superconductivity. 
Therefore, this progress in ferromagnetic semiconductors is a breakthrough in semiconductor physics involving unconventional superconducting pairing.
\end{abstract}
\maketitle


\newcommand{\IFA}{(In,~Fe)As}

Ferromagnetic semiconductors (FMSs) based on III-V materials serve as a link between semiconductor physics and magnetism, and they have exhibited a number of novel phenomena\cite{RevModPhys.86.187} since the discovery of a synthetic method for their fabrication\cite{PhysRevLett.63.1849,:/content/aip/journal/apl/69/3/10.1063/1.118061,HAYASHI19971063}. 
Their application to spintronics\cite{RevModPhys.76.323} has also been seriously considered and many innovative device actions have been found in them, although the origin of ferromagnetism remains under debate\cite{:/content/aip/journal/apr2/1/1/10.1063/1.4840136}.
Indium-based narrow gap semiconductors, however, offer another critical link between semiconductor physics and superconductivity\cite{Schapers201310}, because of the typically low contact resistance between the semiconductors and metals.
Such desirable contact properties are the result of natural band-bending at the surfaces of In-based semiconductors. 
Hence, these properties are restricted to $n$-type semiconductors.
Nevertheless, most of the magnetic ions to date that are used to provide magnetic moments in FMSs work as acceptors, thus turning the host material into $p$-type, rendering the contacts poor.
The exploration of the semiconductor-magnetism-superconductivity crossover in physics, therefore, awaits the detection of an $n$-type narrow-gap FMS, which has been realised as \IFA, in which Be dopants work as double-donors\cite{:/content/aip/journal/apl/101/18/10.1063/1.4764947}.
Later, it was found that \IFA can be $n$-type even without the Be donors depending on its structure and growth condition.

A fundamental topic of interest in superconductivity observed in ferromagnetic materials is the mechanisms of minimising the competition between the superconducting pair potential and the ferromagnetic exchange potential.
Conventional superconductivity is based on Cooper pairs formed by electrons with (momentum,~spin) = ($\hbar\bm{k}, \uparrow$) and ($-\hbar\bm{k}, \downarrow$).
The ferromagnetic exchange potential should be compensated through the modification of their momenta or spins.
In the former, the pairs have non-zero centre-of-mass momentum, forming the so-called Fulde-Ferrell-Larkin-Ovchinnikov (FFLO) state.
In the latter, the spin-triplet pairing takes place and results in a spin-polarized supercurrent, which is, in a sense, a super-spin current.
Such unconventional superconductivity does not only originate from material-specific properties in bulk, but is also realised in hybrid systems. For example, the observation of an FFLO state was reported in a superconductor/ferromagnet/superconductor (SFS) junction\cite{PhysRevB.82.100501}, 
in addition to the observation of the possible realisation of spin-triplet superconductivity in a superconductor/half-metal/superconductor (SHMS) junction\cite{keizer2006spin}.
Many of the III-V FMSs are presumed to be half-metallic, as interpreted from band calculations\cite{PhysRevLett.81.3002,PhysRevB.63.195205}, some of which are experimentally supported by certain phenomena, {\it e.g.} tunnelling magnetoresistance\cite{PhysRevLett.87.026602}.
Therefore, questions such as whether a supercurrent flows in an FMS, and if it does, which type of pairing is realised, are of significant interest both in spintronics and superconductivity.

\begin{figure*}
\includegraphics[width=\linewidth]{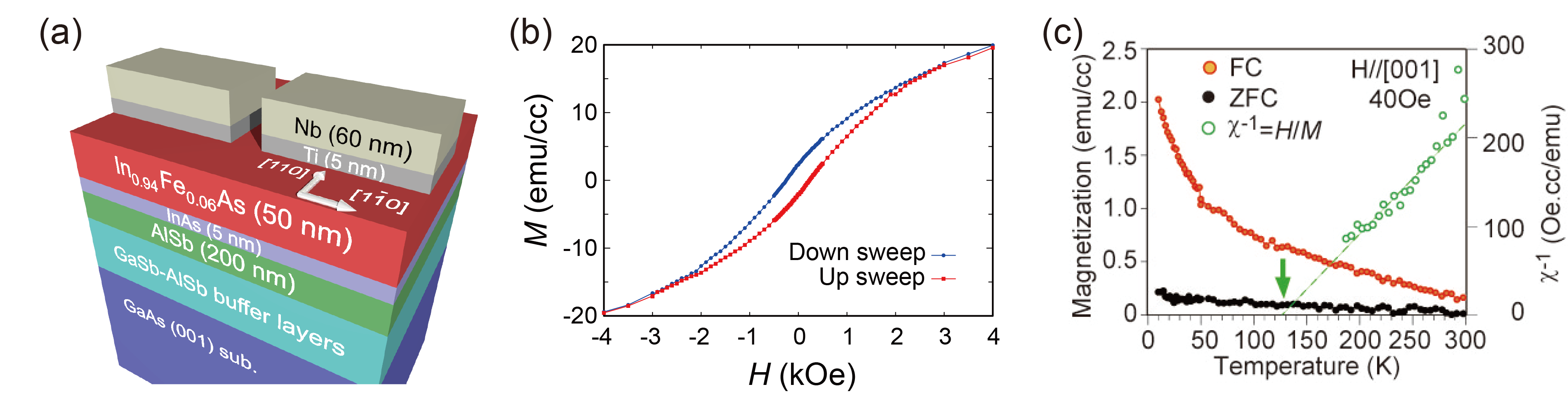}
\caption{\label{fig_sample}
(a) Schematic illustration of the layered structure. 
(b) $M$ (magnetisation) -$H$ (magnetic field) curve of the \IFA\ film at 10~K measured with a dc-SQUID magnetometer (Quantum Design, MPMS). The field sweep range is $\pm10$~kOe.
(c) Temperature dependence of the magnetisation in \IFA\ at $H$=40~Oe for field cooling (FC, $H= 10$~kOe) and zero-field cooling (ZFC). 
The green open circles represent inverse magnetic susceptibility indicating a Curie temperature of 128~K.
}
\end{figure*}

\begin{figure}[b]
\includegraphics[width=\linewidth]{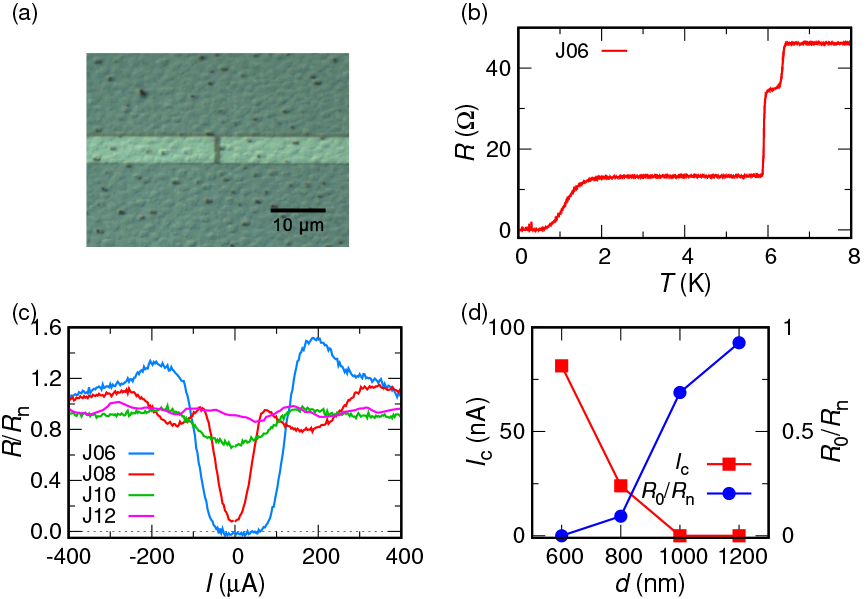}
\caption{\label{fig_ZeroField}
\\
(a) Optical micrograph image of the junction J10.
(b) Temperature dependence of the zero-bias resistance of J06 at zero magnetic fields.
(c) Differential resistance as a function of the bias current for each junction at zero fields.
(d) Dependence of critical current and ratio of the zero bias resistance $R_0$ to the normal resistance $R_\mathrm{n}$ on the gap length. 
}
\end{figure}

The devices used are lateral-type junctions, in which the superconducting electrodes were deposited on top of the \IFA\ heterostructure, as illustrated in Fig.\ref{fig_sample}(a).
Figure \ref{fig_sample}(b) shows the magnetisation curve of the present \IFA\ film at 10~K, which exhibits clear hysteresis owing to ferromagnetism.
The Curie temperature was estimated to be approximately 128~K from the temperature dependence of magnetic susceptibility, as shown in Fig.\ref{fig_sample}(c).
An optical micrograph image of a gap between the electrodes, which we call the ``junction'' henceforth, is shown in \ref{fig_ZeroField}(a). The electric current direction was taken along [$\bar{1}$10] of the \IFA\ crystal. This direction is optimal for generating a supercurrent in \IFA, as examined previously\cite{1742-6596-969-1-012036}
In this experiment we prepared four junctions with gap lengths of 0.6, 0.8, 1.0, and 1.2~$\mu$m, which are named J06, J08, J10, and J12, respectively.

\newcommand{\Ic}{I_\mathrm{c}}

The Nb/Ti electrodes underwent zero-resistance transition around 6~K ($T_\mathrm{c}$), which corresponds to a superconducting gap $\Delta_0$ of 1.0~meV\cite{PhysRevB.57.14416}.
Figure \ref{fig_ZeroField}(b) shows the temperature dependence of the zero-bias, zero-magnetic field resistance in J06.
Below the $T_\mathrm{c}$ of Nb, the resistance stayed constant down to 2~K, under which it started decreasing again with decreasing temperature.
The stepwise temperature variation indicates that the Nb/\IFA\ interfacial resistance dominates the total resistance in the intermediate temperature region and at around 2~K the superconducting proximity areas extending from the Nb electrodes begin to overlap.
At the lowest temperature $\sim 0.1$~K, all the junctions show non-linear $I$-$V$ characteristics, namely a dip structure in d$V$/d$I$ around the zero-bias current, as shown in Fig.\ref{fig_ZeroField}(c), and J06 exhibits clear zero resistance.
To quantify the rather rounded rise of the resistance, we define the critical current $\Ic$ as the current at which the resistance recovers to 20\% of the normal resistance $R_\mathrm{n}$.
Figure \ref{fig_ZeroField}(d) shows the distance dependence of the critical current and the zero-bias resistance.
The critical current decreases with increasing distance, indicating that the superconductivity is not bulk but the proximity effect.
The proximity length is of an order similar to that in triplet proximity systems, as we discuss below\cite{keizer2006spin,Robinson59}.

\begin{figure}[b]
\includegraphics[width=\linewidth]{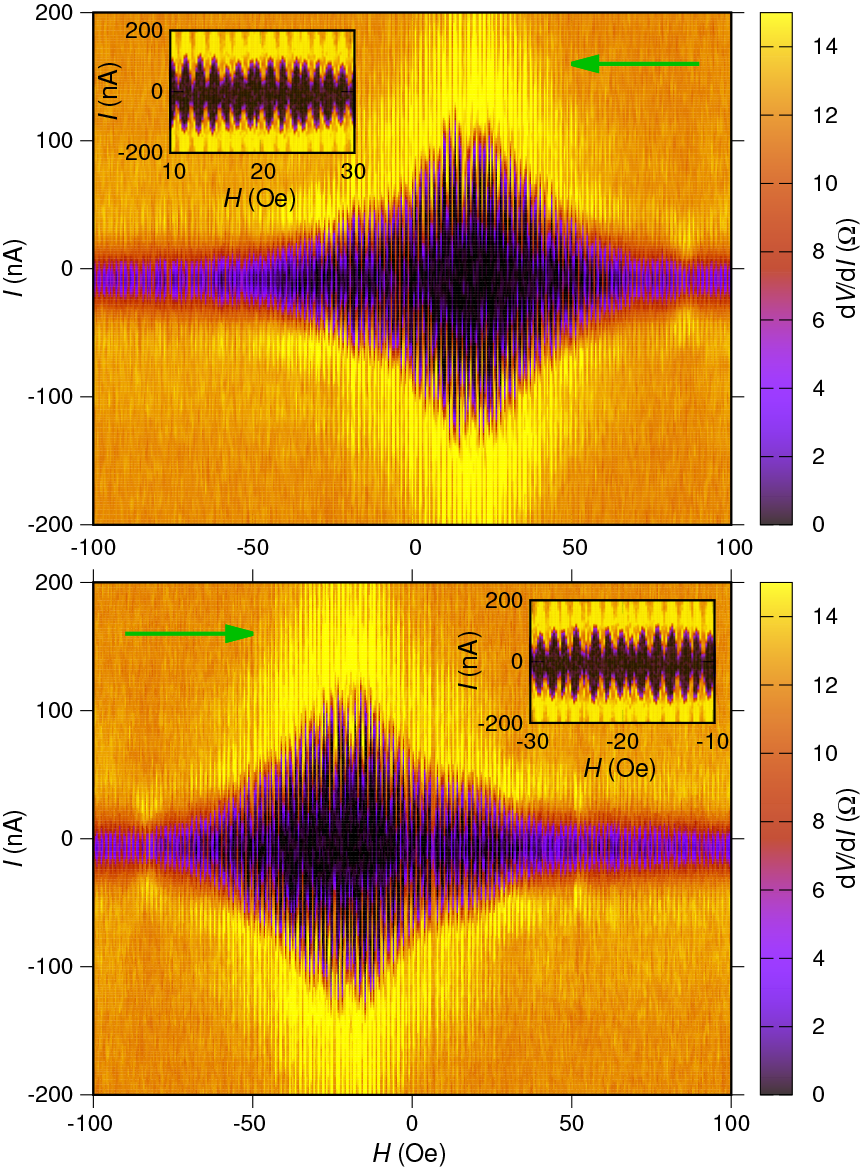}
\caption{\label{fig_Field-dep}
Colour plot of the differential resistance of J06 as a function of magnetic field and bias current. The external field was swept from +109~Oe to -109~Oe for the upper panel, and opposite for the lower panel, in which sweep directions are indicated by the green arrows. The insets are the expansion around +20~Oe and -20~Oe for down-sweep and up-sweep, respectively.
}
\end{figure}

The magnetic field and bias current dependence of the differential resistance in J06 are shown in Fig.\ref{fig_Field-dep}. 
We observe fine regular oscillations in the differential resistance and the critical current against the perpendicular field.
Each zero-resistance region in the oscillation is diamond-shaped on the $H$ (magnetic field) -$I$ (current) plane as shown in the insets of Fig.\ref{fig_Field-dep} or in Fig.5(a).
This oscillatory field-dependence evidences that the supercurrent originates from the Josephson effect.
The observed period of 1.45~Oe is, however, shorter than the simple estimation of 5.17~Oe, corresponding to a single flux quantum $\Phi_0 = h/2e$ per the junction area $A=(d+2\lambda)w=4~\mu$m$^2$, where the gap length $d=600 $~nm, width of the electrodes $w=5~\mu$m, and the penetration depth of Nb $\lambda = 100$~nm\cite{PhysRevB.72.064503}.
This short period is attributable to the flux focusing effect due to the lateral junction configuration and diamagnetism of the Nb electrodes\cite{:/content/aip/journal/apl/59/26/10.1063/1.105660}. 
Considering the flux focusing, the effective junction area $A_\mathrm{eff}$ is given by $f(t,d,w,\lambda)w^2$, where $t$ is the thickness of Nb and a function $f(t,d,w,\lambda)$ is approximated by a constant 0.543 in the thin film region $t/\lambda <2$.
Under the present parameters, the focusing effect boosts the effective junction area to 13.6~$\mu\mathrm{m}^2$, and diminishes the period to 1.52~Oe, which is in reasonable agreement with the experiment.
Even though the junction resistance is no longer zero above 100~Oe, the differential resistance still oscillates with the same period, which confirms the existence of the superconducting coherence via the junction.
Though another junction J08 does not reach zero resistance at zero field, it also exhibits similar resistance oscillations against perpendicular fields.

Still the behaviour in Fig.\ref{fig_Field-dep} is anomalous as an interference pattern in a single Josephson junction.
In an ordinary Fraunhofer pattern, $\Ic$ takes maximum at the flux density $B_0=0$~G for a so-called 0-junction, or maxima at $B_\pi=\pm \Phi_0/(2A_\mathrm{eff}) = 0.76$~G for a $\pi$-junction, and decreases rapidly with increasing magnetic field. The curve of $\Ic$ vs. $B$ is symmetric with respect to the origin $(B, I)=(0,0)$ and independent of the field sweep direction. 
The anomalies can be summarised in the following two points: 
Firstly, damping of the oscillation with the magnetic field is surprisingly weak and the oscillation is observable up to 100~cycles.
Secondly, $\Ic$ becomes maximum much before the flux density $B=\mu_0H+M$ ($M$ : the magnetisation) goes down to zero, as explained below.
The $\Ic$s as a function of $B$ for up and down sweeps are each highly asymmetric with respect to the origin, and furthermore, they are hysteretic for the field. 
This indicates that the observed Josephson effect breaks time-reversal symmetry, clearly reflecting the ferromagnetism in \IFA.
The envelope of the curve of $\Ic$ vs. $B$ for up sweep is mirror symmetric to that for down sweep about $H=0$ when the sweep ranges are centred at $H=0$ and after the field is swept a few times in the same range.
In combination with the magnetisation curve for a wide field range in Fig.\ref{fig_sample}(b),
the above facts indicate that the $M$-$H$ curve in these narrow-range sweeps also has counter-clockwise
loops around the origin of the $M$-$H$ plane, probably due to the small coercive field and repeating field sweeps in the range.
This means, for example, that $M$ is kept positive for down sweeps in the region of $H>0$.
On the other hand, the envelope of $\Ic$ in the upper panel of Fig.\ref{fig_Field-dep} takes the peak around $+20$~Oe for the down sweep, at which field the flux density $B=\mu_0H+M$ should be finite and positive.
In the same way, we know that the peak for the up sweep is around $-20$~Oe, at which $B$ should be finite and negative.

Now we look for a possible explanation of the above observations.
Clear oscillations in the differential resistance versus magnetic field are observable in devices J06 and J08. That is, the superconducting coherence survives up to 0.8~$\mu$m in \IFA\cite{Schapers201310,Irie2014PRB}.
The conventional spin-singlet pairs, however, should be destroyed immediately away from the interface by the ferromagnetic exchange interaction\cite{PhysRevLett.89.137007,doi:10.1063/1.2356104,PhysRevB.74.140501}.
The decay length of the spin-singlet order parameters in ferromagnets $\xi_\mathrm{d}$ is written as follows:
\begin{align}
\xi_\mathrm{d} =
\sqrt{\dfrac{\hbar D}{\sqrt{(\pi k_\mathrm{B}T)^2 + E_\mathrm{ex}^2} + \pi k_\mathrm{B}T}},
\end{align}
where $D$ and $E_\mathrm{ex}$ are the diffusion coefficient and exchange field in ferromagnets, respectively\cite{BuzdinJETP1992}. 
Anh \textit{et al}. demonstrated that the exchange energies of \IFA\ were well-explained by the Brillouin function\cite{Anh2016Ncomm}. 
As a result of their research, we can approximate the exchange energy on this basis of Curie's law and the Brillouin function.
Therefore we estimate the exchange energy $E_\mathrm{ex}$ in the present \IFA\ to be 98~meV from the Curie temperature.
Using $D=1.4\times10^{-3}~\mathrm{m^2/s}$ in the present \IFA, the decay length $\xi_\mathrm{d}$ is estimated to be 3.1~nm at 0.1~K.
The gap lengths of our devices are much longer than the distance by which the spin-singlet order parameters from the Nb electrodes are expected to overlap with each other, even if we consider an empirical rule that we can observe the Josephson effect in junctions with gap lengths approximately ten times greater than $\xi_\mathrm{d}$.
Because the present \IFA\ film is undoubtedly ferromagnetic, the present pairing in the proximity-induced superconductivity must be the spin-triplet from the above discussion.

Next we consider the interference patterns observed in Fig.\ref{fig_Field-dep}. 
As the first point, the weak damping in amplitude and the regularity in the period of oscillations indicates that the current density is strongly localised at the two edges of the junction area as calculated, {\it e.g.} in refs.\cite{doi:10.1002/pssa.2210410206,BaronePaterno198205}.
This is, however, reasonable if we consider the Josephson penetration depth $\lambda_{\rm J}$.
In an ordinary expression, it is $(\hbar/2e\mu_0d\Ic)^{1/2}$, where $d$ is the junction width and very large now, but here we need to replace
$\mu_0$ with some ``effective" permittivity, which contains the effect of the ferromagnet and the flux-concentration by superconducting electrodes, and hence, is also very large. Consequently, $\lambda_{\rm J}$ must be very small, leading to the strong current localisation at the edges.

\begin{figure}[b]
\includegraphics[width=\linewidth]{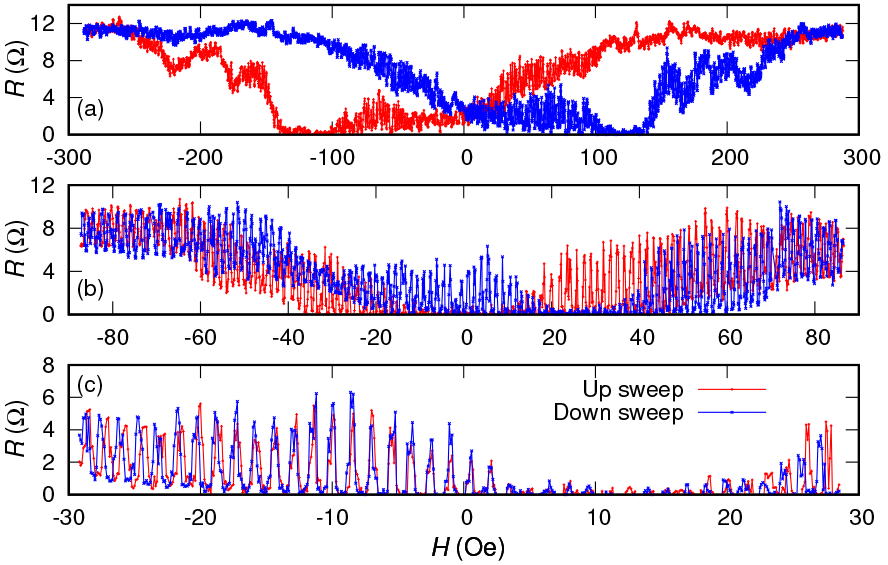}
\caption{\label{fig_ZBR-H}
Field dependence of the zero-bias differential resistance of J06. The field was swept sequentially via the following points: +14~kOe, -290, 290, -290, 290, -90, 90, -90, 90, -30, 30, -30, and 30~Oe. The panels are the data for the sweeps (a) from -290~Oe to -290~Oe via +290~Oe, (b) from -90~Oe to -90~Oe via +90~Oe, and (c) from -30~Oe to -30~Oe via +30~Oe, respectively.}
\end{figure}

The second point---the peak positions of the envelope---gives interesting information.
Inside the film, the flux density is still $\mu_0H+M$, while the magnetic field is $H_{\rm in}=H+NM/\mu_0$, where $N$, the demagnetisation coefficient,
is almost $-1$ for thin films. Assuming an $M$-$H$ curve similar to that in Fig.\ref{fig_sample}(b) for the minor loop in Fig.\ref{fig_Field-dep},
and considering the flux-concentration effect, we identify the peak positions corresponding to $H_{\rm in}\sim 0$.
In superconductivity with singlet pairing, there is no such electromagnetic freedom that picks up the local magnetic field, though the spin of Cooper pairs can do that in the triplet pairing superconductivity.
More specifically, because of the granularity in the ferromagnetism, 
the randomness in the magnetisation inside the film should become maximum at $H_{\rm in}=0$,
and this could be the best condition for the triplet proximity effect from singlet superconductors\cite{PhysRevLett.110.237003}.

The granularity of the ferromagnetism also appears in the minor loop behaviour.
In Fig.\ref{fig_ZBR-H}, we plot the field dependence of the zero-bias differential resistance (ZBR) in different sweep ranges.
The ZBR curves are hysteretic when the sweep range exceeds 90~Oe as shown in Fig.\ref{fig_ZBR-H}(a)(b), but no hysteresis is observable for the sweep range narrower than 30~Oe, as shown in Fig.\ref{fig_ZBR-H}(c).
This behaviour reflects the granular ferromagnetism in \IFA\ and some ratchet-like mechanisms which prevent instantaneous reversal of domain wall motion and are initiated between 30 and 90~Oe.

\begin{figure}
\includegraphics[width=\linewidth]{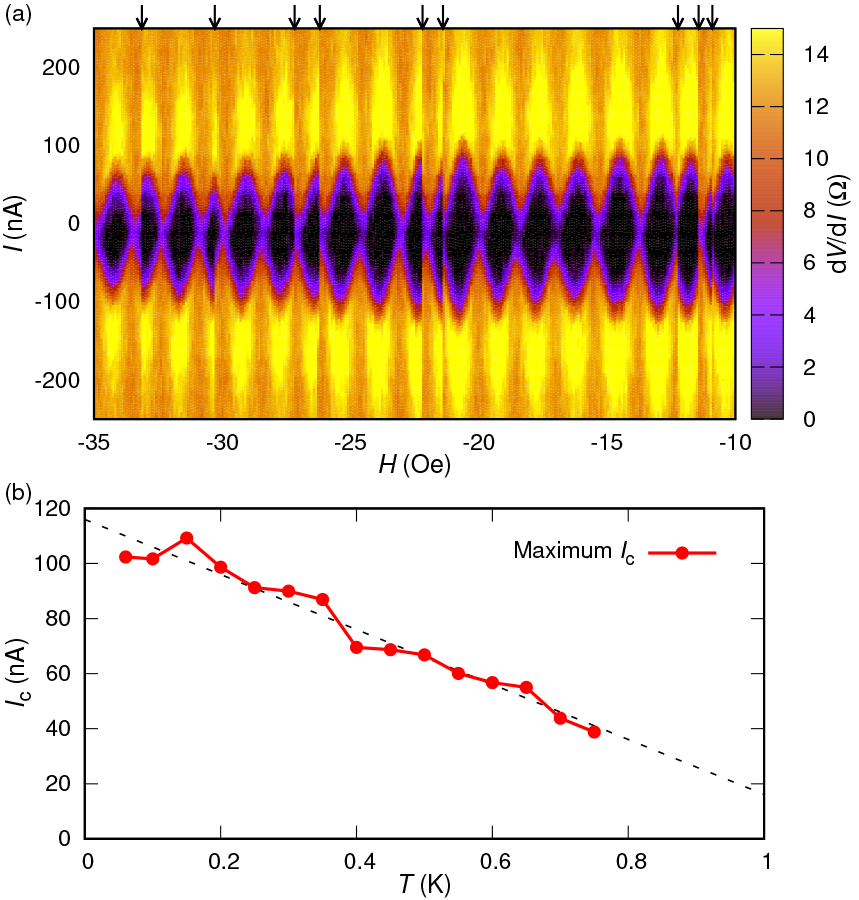}
\caption{\label{fig_higherT}
(a) Colour plot of the differential resistance of J06 as a function of the magnetic field and bias current at 0.35~K. After the field was swept to -109~Oe, the resistance measured from -36~Oe to -8~Oe. The arrows above the graph indicate the phase slip attributable to the domain wall motion.
(b) Temperature dependence of the maximum critical current of J06 in the range between -36~Oe and -8~Oe. The broken line is a fitting line of the data.}
\end{figure}

Such domain motion sometimes contains jumps in fluxoid at the junction,
which are observable in the oscillation pattern in Fig.\ref{fig_higherT}(a). 
In spite of such jumps, the envelope of $\Ic$ does not change in accordance with our interpretation that 
the interference pattern depends on the magnetic flux piercing the junction area while the current is localised at the edges and the amplitude is determined by the condition of singlet-triplet connection.
The emergence of finite resistance below 0.5~K shown in Fig.\ref{fig_ZeroField}(b) is also caused by this discontinuous phase slip, indicating domain wall motion by the temperature sweep. 
This makes it difficult to fix the phase difference during the measurement of the temperature dependence of $\Ic$.
To avoid this difficulty,
we performed the same measurement as in Fig.\ref{fig_higherT}(a) and obtained the maximum $\Ic$ in a certain field range at various temperatures.
Figure \ref{fig_higherT}(b) is the temperature dependence of the maximum $\Ic$ thus obtained, which monotonically decreases with increasing temperature in the same manner as in previous works\cite{keizer2006spin}.

Thus far we have seen that the spin-triplet pairing scenario can explain all the observations, 
whereas the conventional spin-singlet picture cannot evade essential difficulties.
At the same time, \IFA\ is expected to satisfy several conditions to induce triplet pairing from singlet superconductors, e.g., inhomogeneous magnetisation, spin-orbit interaction, and high spin polarization.
The inhomogeneous magnetisation, which causes spin scattering matrices generating triplet component at the surface\cite{NatPhys.4.138}, is naturally formed at the surface of III-V ferromagnetic semiconductors due to the strain-induced magnetisation-reorientation\cite{RevModPhys.86.187}.
In addition, strong spin-orbit interaction exists in \IFA\ as a narrow-gap semiconductor.
Although the spin-orbit interaction inevitably leads to some mixing of singlet-triplet superconductivity, only the triplet component survives in \IFA\ due to its ferromagnetic exchange interaction\cite{Anh2016Ncomm}.
Note that the present \IFA\ film has a fairly short mean-free path and is close to the dirty limit. From the study of the impurity effect in noncentrosymmetric superconductors, it has been clarified that, even when the spin-orbit interaction is finite, the $s$-wave coupling can be dominant. 
We then conclude that the $s$-wave spin-triplet pairing, which was observed in the S/HM/S junctions\cite{keizer2006spin}, is dominant in the observed superconducting proximity effect.

\section{Methods}
The 50-nm-thick \IFA\ film was grown by low-temperature molecular beam epitaxy on a (001) GaAs substrate with AlSb-based buffer layers, as displayed in Fig.\ref{fig_sample}(a).
Details of the growth are given elsewhere\cite{:/content/aip/journal/apl/101/18/10.1063/1.4764947}.
The Fe concentration was 6\% and the carrier concentration was estimated to be 8$\times 10^{18}~\mathrm{cm}^{-3}$ at 3.5~K without the Be donor. 
The mean free path was also estimated to be approximately 4 nm, shorter than the thickness of the \IFA\ film. 
After fabrication of the lift-off pattern with electron-beam lithography, the Ti/Nb electrodes were deposited with ion-beam sputtering, immediately after the surface cleaning with Ar ion beam.
The specimens were cooled down in a dilution fridge. Each electric connection had a low-pass filter heat-anchored to the mixing chamber. The differential resistance $R=$d$V$/d$I$ of the junctions was measured by using a lock-in technique with an ac current modulation of 5~nA-rms at a frequency of 89 or 890~Hz superimposed on the dc current.

\section{Acknowledgements}
We wish to thank Y. Asano for fruitful discussions.
This work was supported by Grants-in-Aid for Scientific Research on Innovative Areas, ``Nano Spin Conversion Science" (Grant No.26103003) and ``Topological Materials Science" (Grant No.18H04218), and
also by JSPS KAKENHI Grant Numbers 25247051, 17K05492, 17H04922, and 18H03860
and by Special Coordination Funds for Promoting Science and Technology.


\begin{thebibliography}{28}%
\makeatletter
\providecommand \@ifxundefined [1]{%
 \@ifx{#1\undefined}
}%
\providecommand \@ifnum [1]{%
 \ifnum #1\expandafter \@firstoftwo
 \else \expandafter \@secondoftwo
 \fi
}%
\providecommand \@ifx [1]{%
 \ifx #1\expandafter \@firstoftwo
 \else \expandafter \@secondoftwo
 \fi
}%
\providecommand \natexlab [1]{#1}%
\providecommand \enquote  [1]{``#1''}%
\providecommand \bibnamefont  [1]{#1}%
\providecommand \bibfnamefont [1]{#1}%
\providecommand \citenamefont [1]{#1}%
\providecommand \href@noop [0]{\@secondoftwo}%
\providecommand \href [0]{\begingroup \@sanitize@url \@href}%
\providecommand \@href[1]{\@@startlink{#1}\@@href}%
\providecommand \@@href[1]{\endgroup#1\@@endlink}%
\providecommand \@sanitize@url [0]{\catcode `\\12\catcode `\$12\catcode
  `\&12\catcode `\#12\catcode `\^12\catcode `\_12\catcode `\%12\relax}%
\providecommand \@@startlink[1]{}%
\providecommand \@@endlink[0]{}%
\providecommand \url  [0]{\begingroup\@sanitize@url \@url }%
\providecommand \@url [1]{\endgroup\@href {#1}{\urlprefix }}%
\providecommand \urlprefix  [0]{URL }%
\providecommand \Eprint [0]{\href }%
\providecommand \doibase [0]{http://dx.doi.org/}%
\providecommand \selectlanguage [0]{\@gobble}%
\providecommand \bibinfo  [0]{\@secondoftwo}%
\providecommand \bibfield  [0]{\@secondoftwo}%
\providecommand \translation [1]{[#1]}%
\providecommand \BibitemOpen [0]{}%
\providecommand \bibitemStop [0]{}%
\providecommand \bibitemNoStop [0]{.\EOS\space}%
\providecommand \EOS [0]{\spacefactor3000\relax}%
\providecommand \BibitemShut  [1]{\csname bibitem#1\endcsname}%
\let\auto@bib@innerbib\@empty
\bibitem [{\citenamefont {Dietl}\ and\ \citenamefont
  {Ohno}(2014)}]{RevModPhys.86.187}%
  \BibitemOpen
  \bibfield  {author} {\bibinfo {author} {\bibfnamefont {T.}~\bibnamefont
  {Dietl}}\ and\ \bibinfo {author} {\bibfnamefont {H.}~\bibnamefont {Ohno}},\
  }\href {\doibase 10.1103/RevModPhys.86.187} {\bibfield  {journal} {\bibinfo
  {journal} {Rev. Mod. Phys.}\ }\textbf {\bibinfo {volume} {86}},\ \bibinfo
  {pages} {187} (\bibinfo {year} {2014})}\BibitemShut {NoStop}%
\bibitem [{\citenamefont {Munekata}\ \emph {et~al.}(1989)\citenamefont
  {Munekata}, \citenamefont {Ohno}, \citenamefont {von Molnar}, \citenamefont
  {Segm\"uller}, \citenamefont {Chang},\ and\ \citenamefont
  {Esaki}}]{PhysRevLett.63.1849}%
  \BibitemOpen
  \bibfield  {author} {\bibinfo {author} {\bibfnamefont {H.}~\bibnamefont
  {Munekata}}, \bibinfo {author} {\bibfnamefont {H.}~\bibnamefont {Ohno}},
  \bibinfo {author} {\bibfnamefont {S.}~\bibnamefont {von Molnar}}, \bibinfo
  {author} {\bibfnamefont {A.}~\bibnamefont {Segm\"uller}}, \bibinfo {author}
  {\bibfnamefont {L.~L.}\ \bibnamefont {Chang}}, \ and\ \bibinfo {author}
  {\bibfnamefont {L.}~\bibnamefont {Esaki}},\ }\href {\doibase
  10.1103/PhysRevLett.63.1849} {\bibfield  {journal} {\bibinfo  {journal}
  {Phys. Rev. Lett.}\ }\textbf {\bibinfo {volume} {63}},\ \bibinfo {pages}
  {1849} (\bibinfo {year} {1989})}\BibitemShut {NoStop}%
\bibitem [{\citenamefont {Ohno}\ \emph {et~al.}(1996)\citenamefont {Ohno},
  \citenamefont {Shen}, \citenamefont {Matsukura}, \citenamefont {Oiwa},
  \citenamefont {Endo}, \citenamefont {Katsumoto},\ and\ \citenamefont
  {Iye}}]{:/content/aip/journal/apl/69/3/10.1063/1.118061}%
  \BibitemOpen
  \bibfield  {author} {\bibinfo {author} {\bibfnamefont {H.}~\bibnamefont
  {Ohno}}, \bibinfo {author} {\bibfnamefont {A.}~\bibnamefont {Shen}}, \bibinfo
  {author} {\bibfnamefont {F.}~\bibnamefont {Matsukura}}, \bibinfo {author}
  {\bibfnamefont {A.}~\bibnamefont {Oiwa}}, \bibinfo {author} {\bibfnamefont
  {A.}~\bibnamefont {Endo}}, \bibinfo {author} {\bibfnamefont {S.}~\bibnamefont
  {Katsumoto}}, \ and\ \bibinfo {author} {\bibfnamefont {Y.}~\bibnamefont
  {Iye}},\ }\href {\doibase http://dx.doi.org/10.1063/1.118061} {\bibfield
  {journal} {\bibinfo  {journal} {Appl. Phys. Lett.}\ }\textbf {\bibinfo
  {volume} {69}},\ \bibinfo {pages} {363} (\bibinfo {year} {1996})}\BibitemShut
  {NoStop}%
\bibitem [{\citenamefont {Hayashi}\ \emph {et~al.}(1997)\citenamefont
  {Hayashi}, \citenamefont {Tanaka}, \citenamefont {Nishinaga}, \citenamefont
  {Shimada}, \citenamefont {Tsuchiya},\ and\ \citenamefont
  {Otuka}}]{HAYASHI19971063}%
  \BibitemOpen
  \bibfield  {author} {\bibinfo {author} {\bibfnamefont {T.}~\bibnamefont
  {Hayashi}}, \bibinfo {author} {\bibfnamefont {M.}~\bibnamefont {Tanaka}},
  \bibinfo {author} {\bibfnamefont {T.}~\bibnamefont {Nishinaga}}, \bibinfo
  {author} {\bibfnamefont {H.}~\bibnamefont {Shimada}}, \bibinfo {author}
  {\bibfnamefont {H.}~\bibnamefont {Tsuchiya}}, \ and\ \bibinfo {author}
  {\bibfnamefont {Y.}~\bibnamefont {Otuka}},\ }\href {\doibase
  http://dx.doi.org/10.1016/S0022-0248(96)00937-2} {\bibfield  {journal}
  {\bibinfo  {journal} {J. Cryst. Growth}\ }\textbf {\bibinfo {volume} {175}},\
  \bibinfo {pages} {1063 } (\bibinfo {year} {1997})}\BibitemShut {NoStop}%
\bibitem [{\citenamefont {\ifmmode \check{Z}\else
  \v{Z}\fi{}uti\ifmmode~\acute{c}\else \'{c}\fi{}}\ \emph
  {et~al.}(2004)\citenamefont {\ifmmode \check{Z}\else
  \v{Z}\fi{}uti\ifmmode~\acute{c}\else \'{c}\fi{}}, \citenamefont {Fabian},\
  and\ \citenamefont {Das~Sarma}}]{RevModPhys.76.323}%
  \BibitemOpen
  \bibfield  {author} {\bibinfo {author} {\bibfnamefont {I.}~\bibnamefont
  {\ifmmode \check{Z}\else \v{Z}\fi{}uti\ifmmode~\acute{c}\else \'{c}\fi{}}},
  \bibinfo {author} {\bibfnamefont {J.}~\bibnamefont {Fabian}}, \ and\ \bibinfo
  {author} {\bibfnamefont {S.}~\bibnamefont {Das~Sarma}},\ }\href {\doibase
  10.1103/RevModPhys.76.323} {\bibfield  {journal} {\bibinfo  {journal} {Rev.
  Mod. Phys.}\ }\textbf {\bibinfo {volume} {76}},\ \bibinfo {pages} {323}
  (\bibinfo {year} {2004})}\BibitemShut {NoStop}%
\bibitem [{\citenamefont {Tanaka}\ \emph {et~al.}(2014)\citenamefont {Tanaka},
  \citenamefont {Ohya},\ and\ \citenamefont
  {Nam~Hai}}]{:/content/aip/journal/apr2/1/1/10.1063/1.4840136}%
  \BibitemOpen
  \bibfield  {author} {\bibinfo {author} {\bibfnamefont {M.}~\bibnamefont
  {Tanaka}}, \bibinfo {author} {\bibfnamefont {S.}~\bibnamefont {Ohya}}, \ and\
  \bibinfo {author} {\bibfnamefont {P.}~\bibnamefont {Nam~Hai}},\ }\href
  {\doibase http://dx.doi.org/10.1063/1.4840136} {\bibfield  {journal}
  {\bibinfo  {journal} {Applied Physics Reviews}\ }\textbf {\bibinfo {volume}
  {1}},\ \bibinfo {eid} {011102} (\bibinfo {year} {2014}),\
  }\BibitemShut {NoStop}%
\bibitem [{\citenamefont {Sch\"apers}(2011)}]{Schapers201310}%
  \BibitemOpen
  \bibfield  {author} {\bibinfo {author} {\bibfnamefont {T.}~\bibnamefont
  {Sch\"apers}},\ }\href {http://amazon.com/o/ASIN/364207586X/} {\emph
  {\bibinfo {title} {Superconductor/Semiconductor Junctions}}}\ (\bibinfo
  {publisher} {Springer, New York},\ \bibinfo {year} {2011})\BibitemShut {NoStop}%
\bibitem [{\citenamefont {Nam~Hai}\ \emph {et~al.}(2012)\citenamefont
  {Nam~Hai}, \citenamefont {Duc~Anh}, \citenamefont {Mohan}, \citenamefont
  {Tamegai}, \citenamefont {Kodzuka}, \citenamefont {Ohkubo}, \citenamefont
  {Hono},\ and\ \citenamefont
  {Tanaka}}]{:/content/aip/journal/apl/101/18/10.1063/1.4764947}%
  \BibitemOpen
  \bibfield  {author} {\bibinfo {author} {\bibfnamefont {P.}~\bibnamefont
  {Nam~Hai}}, \bibinfo {author} {\bibfnamefont {L.}~\bibnamefont {Duc~Anh}},
  \bibinfo {author} {\bibfnamefont {S.}~\bibnamefont {Mohan}}, \bibinfo
  {author} {\bibfnamefont {T.}~\bibnamefont {Tamegai}}, \bibinfo {author}
  {\bibfnamefont {M.}~\bibnamefont {Kodzuka}}, \bibinfo {author} {\bibfnamefont
  {T.}~\bibnamefont {Ohkubo}}, \bibinfo {author} {\bibfnamefont
  {K.}~\bibnamefont {Hono}}, \ and\ \bibinfo {author} {\bibfnamefont
  {M.}~\bibnamefont {Tanaka}},\ }\href {\doibase
  http://dx.doi.org/10.1063/1.4764947} {\bibfield  {journal} {\bibinfo
  {journal} {Appl. Phys. Lett.}\ }\textbf {\bibinfo {volume} {101}},\ \bibinfo
  {eid} {182403} (\bibinfo {year} {2012}),\
  }\BibitemShut {NoStop}%
\bibitem [{\citenamefont {Anwar}\ \emph {et~al.}(2010)\citenamefont {Anwar},
  \citenamefont {Czeschka}, \citenamefont {Hesselberth}, \citenamefont
  {Porcu},\ and\ \citenamefont {Aarts}}]{PhysRevB.82.100501}%
  \BibitemOpen
  \bibfield  {author} {\bibinfo {author} {\bibfnamefont {M.~S.}\ \bibnamefont
  {Anwar}}, \bibinfo {author} {\bibfnamefont {F.}~\bibnamefont {Czeschka}},
  \bibinfo {author} {\bibfnamefont {M.}~\bibnamefont {Hesselberth}}, \bibinfo
  {author} {\bibfnamefont {M.}~\bibnamefont {Porcu}}, \ and\ \bibinfo {author}
  {\bibfnamefont {J.}~\bibnamefont {Aarts}},\ }\href {\doibase
  10.1103/PhysRevB.82.100501} {\bibfield  {journal} {\bibinfo  {journal} {Phys.
  Rev. B}\ }\textbf {\bibinfo {volume} {82}},\ \bibinfo {pages} {100501}
  (\bibinfo {year} {2010})}\BibitemShut {NoStop}%
\bibitem [{\citenamefont {Keizer}\ \emph {et~al.}(2006)\citenamefont {Keizer},
  \citenamefont {Goennenwein}, \citenamefont {Klapwijk}, \citenamefont {Miao},
  \citenamefont {Xiao},\ and\ \citenamefont {Gupta}}]{keizer2006spin}%
  \BibitemOpen
  \bibfield  {author} {\bibinfo {author} {\bibfnamefont {R.}~\bibnamefont
  {Keizer}}, \bibinfo {author} {\bibfnamefont {S.}~\bibnamefont {Goennenwein}},
  \bibinfo {author} {\bibfnamefont {T.}~\bibnamefont {Klapwijk}}, \bibinfo
  {author} {\bibfnamefont {G.}~\bibnamefont {Miao}}, \bibinfo {author}
  {\bibfnamefont {G.}~\bibnamefont {Xiao}}, \ and\ \bibinfo {author}
  {\bibfnamefont {A.}~\bibnamefont {Gupta}},\ }\href@noop {} {\bibfield
  {journal} {\bibinfo  {journal} {Nature}\ }\textbf {\bibinfo {volume} {439}},\
  \bibinfo {pages} {825} (\bibinfo {year} {2006})}\BibitemShut {NoStop}%
\bibitem [{\citenamefont {Akai}(1998)}]{PhysRevLett.81.3002}%
  \BibitemOpen
  \bibfield  {author} {\bibinfo {author} {\bibfnamefont {H.}~\bibnamefont
  {Akai}},\ }\href {\doibase 10.1103/PhysRevLett.81.3002} {\bibfield  {journal}
  {\bibinfo  {journal} {Phys. Rev. Lett.}\ }\textbf {\bibinfo {volume} {81}},\
  \bibinfo {pages} {3002} (\bibinfo {year} {1998})}\BibitemShut {NoStop}%
\bibitem [{\citenamefont {Dietl}\ \emph {et~al.}(2001)\citenamefont {Dietl},
  \citenamefont {Ohno},\ and\ \citenamefont {Matsukura}}]{PhysRevB.63.195205}%
  \BibitemOpen
  \bibfield  {author} {\bibinfo {author} {\bibfnamefont {T.}~\bibnamefont
  {Dietl}}, \bibinfo {author} {\bibfnamefont {H.}~\bibnamefont {Ohno}}, \ and\
  \bibinfo {author} {\bibfnamefont {F.}~\bibnamefont {Matsukura}},\ }\href
  {\doibase 10.1103/PhysRevB.63.195205} {\bibfield  {journal} {\bibinfo
  {journal} {Phys. Rev. B}\ }\textbf {\bibinfo {volume} {63}},\ \bibinfo
  {pages} {195205} (\bibinfo {year} {2001})}\BibitemShut {NoStop}%
\bibitem [{\citenamefont {Tanaka}\ and\ \citenamefont
  {Higo}(2001)}]{PhysRevLett.87.026602}%
  \BibitemOpen
  \bibfield  {author} {\bibinfo {author} {\bibfnamefont {M.}~\bibnamefont
  {Tanaka}}\ and\ \bibinfo {author} {\bibfnamefont {Y.}~\bibnamefont {Higo}},\
  }\href {\doibase 10.1103/PhysRevLett.87.026602} {\bibfield  {journal}
  {\bibinfo  {journal} {Phys. Rev. Lett.}\ }\textbf {\bibinfo {volume} {87}},\
  \bibinfo {pages} {026602} (\bibinfo {year} {2001})}\BibitemShut {NoStop}%
\bibitem [{\citenamefont {Nakamura}\ \emph {et~al.}(2018)\citenamefont
  {Nakamura}, \citenamefont {Anh}, \citenamefont {Hashimoto}, \citenamefont
  {Iwasaki}, \citenamefont {Ohya}, \citenamefont {Tanaka},\ and\ \citenamefont
  {Katsumoto}}]{1742-6596-969-1-012036}%
  \BibitemOpen
  \bibfield  {author} {\bibinfo {author} {\bibfnamefont {T.}~\bibnamefont
  {Nakamura}}, \bibinfo {author} {\bibfnamefont {L.~D.}\ \bibnamefont {Anh}},
  \bibinfo {author} {\bibfnamefont {Y.}~\bibnamefont {Hashimoto}}, \bibinfo
  {author} {\bibfnamefont {Y.}~\bibnamefont {Iwasaki}}, \bibinfo {author}
  {\bibfnamefont {S.}~\bibnamefont {Ohya}}, \bibinfo {author} {\bibfnamefont
  {M.}~\bibnamefont {Tanaka}}, \ and\ \bibinfo {author} {\bibfnamefont
  {S.}~\bibnamefont {Katsumoto}},\ }\href
  {http://stacks.iop.org/1742-6596/969/i=1/a=012036} {\bibfield  {journal}
  {\bibinfo  {journal} {Journal of Physics: Conference Series}\ }\textbf
  {\bibinfo {volume} {969}},\ \bibinfo {pages} {012036} (\bibinfo {year}
  {2018})}\BibitemShut {NoStop}%
\bibitem [{\citenamefont {Pronin}\ \emph {et~al.}(1998)\citenamefont {Pronin},
  \citenamefont {Dressel}, \citenamefont {Pimenov}, \citenamefont {Loidl},
  \citenamefont {Roshchin},\ and\ \citenamefont {Greene}}]{PhysRevB.57.14416}%
  \BibitemOpen
  \bibfield  {author} {\bibinfo {author} {\bibfnamefont {A.~V.}\ \bibnamefont
  {Pronin}}, \bibinfo {author} {\bibfnamefont {M.}~\bibnamefont {Dressel}},
  \bibinfo {author} {\bibfnamefont {A.}~\bibnamefont {Pimenov}}, \bibinfo
  {author} {\bibfnamefont {A.}~\bibnamefont {Loidl}}, \bibinfo {author}
  {\bibfnamefont {I.~V.}\ \bibnamefont {Roshchin}}, \ and\ \bibinfo {author}
  {\bibfnamefont {L.~H.}\ \bibnamefont {Greene}},\ }\href {\doibase
  10.1103/PhysRevB.57.14416} {\bibfield  {journal} {\bibinfo  {journal} {Phys.
  Rev. B}\ }\textbf {\bibinfo {volume} {57}},\ \bibinfo {pages} {14416}
  (\bibinfo {year} {1998})}\BibitemShut {NoStop}%
\bibitem [{\citenamefont {Robinson}\ \emph {et~al.}(2010)\citenamefont
  {Robinson}, \citenamefont {Witt},\ and\ \citenamefont
  {Blamire}}]{Robinson59}%
  \BibitemOpen
  \bibfield  {author} {\bibinfo {author} {\bibfnamefont {J.~W.~A.}\
  \bibnamefont {Robinson}}, \bibinfo {author} {\bibfnamefont {J.~D.~S.}\
  \bibnamefont {Witt}}, \ and\ \bibinfo {author} {\bibfnamefont {M.~G.}\
  \bibnamefont {Blamire}},\ }\href {\doibase 10.1126/science.1189246}
  {\bibfield  {journal} {\bibinfo  {journal} {Science}\ }\textbf {\bibinfo
  {volume} {329}},\ \bibinfo {pages} {59} (\bibinfo {year} {2010})},\ \Eprint
  {http://science.sciencemag.org/content/329/5987/59.full.pdf} \BibitemShut
  {NoStop}%
\bibitem [{\citenamefont {Gubin}\ \emph {et~al.}(2005)\citenamefont {Gubin},
  \citenamefont {Il'in}, \citenamefont {Vitusevich}, \citenamefont {Siegel},\
  and\ \citenamefont {Klein}}]{PhysRevB.72.064503}%
  \BibitemOpen
  \bibfield  {author} {\bibinfo {author} {\bibfnamefont {A.~I.}\ \bibnamefont
  {Gubin}}, \bibinfo {author} {\bibfnamefont {K.~S.}\ \bibnamefont {Il'in}},
  \bibinfo {author} {\bibfnamefont {S.~A.}\ \bibnamefont {Vitusevich}},
  \bibinfo {author} {\bibfnamefont {M.}~\bibnamefont {Siegel}}, \ and\ \bibinfo
  {author} {\bibfnamefont {N.}~\bibnamefont {Klein}},\ }\href {\doibase
  10.1103/PhysRevB.72.064503} {\bibfield  {journal} {\bibinfo  {journal} {Phys.
  Rev. B}\ }\textbf {\bibinfo {volume} {72}},\ \bibinfo {pages} {064503}
  (\bibinfo {year} {2005})}\BibitemShut {NoStop}%
\bibitem [{\citenamefont {Rosenthal}\ \emph {et~al.}(1991)\citenamefont
  {Rosenthal}, \citenamefont {Beasley}, \citenamefont {Char}, \citenamefont
  {Colclough},\ and\ \citenamefont
  {Zaharchuk}}]{:/content/aip/journal/apl/59/26/10.1063/1.105660}%
  \BibitemOpen
  \bibfield  {author} {\bibinfo {author} {\bibfnamefont {P.~A.}\ \bibnamefont
  {Rosenthal}}, \bibinfo {author} {\bibfnamefont {M.~R.}\ \bibnamefont
  {Beasley}}, \bibinfo {author} {\bibfnamefont {K.}~\bibnamefont {Char}},
  \bibinfo {author} {\bibfnamefont {M.~S.}\ \bibnamefont {Colclough}}, \ and\
  \bibinfo {author} {\bibfnamefont {G.}~\bibnamefont {Zaharchuk}},\ }\href
  {\doibase http://dx.doi.org/10.1063/1.105660} {\bibfield  {journal} {\bibinfo
   {journal} {Applied Physics Letters}\ }\textbf {\bibinfo {volume} {59}},\
  \bibinfo {pages} {3482} (\bibinfo {year} {1991})}\BibitemShut {NoStop}%
\bibitem [{\citenamefont {Irie}\ \emph {et~al.}(2014)\citenamefont {Irie},
  \citenamefont {Harada}, \citenamefont {Sugiyama},\ and\ \citenamefont
  {Akazaki}}]{Irie2014PRB}%
  \BibitemOpen
  \bibfield  {author} {\bibinfo {author} {\bibfnamefont {H.}~\bibnamefont
  {Irie}}, \bibinfo {author} {\bibfnamefont {Y.}~\bibnamefont {Harada}},
  \bibinfo {author} {\bibfnamefont {H.}~\bibnamefont {Sugiyama}}, \ and\
  \bibinfo {author} {\bibfnamefont {T.}~\bibnamefont {Akazaki}},\ }\href
  {\doibase 10.1103/PhysRevB.89.165415} {\bibfield  {journal} {\bibinfo
  {journal} {Phys. Rev. B}\ }\textbf {\bibinfo {volume} {89}},\ \bibinfo
  {pages} {165415} (\bibinfo {year} {2014})}\BibitemShut {NoStop}%
\bibitem [{\citenamefont {Kontos}\ \emph {et~al.}(2002)\citenamefont {Kontos},
  \citenamefont {Aprili}, \citenamefont {Lesueur}, \citenamefont {Gen\^et},
  \citenamefont {Stephanidis},\ and\ \citenamefont
  {Boursier}}]{PhysRevLett.89.137007}%
  \BibitemOpen
  \bibfield  {author} {\bibinfo {author} {\bibfnamefont {T.}~\bibnamefont
  {Kontos}}, \bibinfo {author} {\bibfnamefont {M.}~\bibnamefont {Aprili}},
  \bibinfo {author} {\bibfnamefont {J.}~\bibnamefont {Lesueur}}, \bibinfo
  {author} {\bibfnamefont {F.}~\bibnamefont {Gen\^et}}, \bibinfo {author}
  {\bibfnamefont {B.}~\bibnamefont {Stephanidis}}, \ and\ \bibinfo {author}
  {\bibfnamefont {R.}~\bibnamefont {Boursier}},\ }\href {\doibase
  10.1103/PhysRevLett.89.137007} {\bibfield  {journal} {\bibinfo  {journal}
  {Phys. Rev. Lett.}\ }\textbf {\bibinfo {volume} {89}},\ \bibinfo {pages}
  {137007} (\bibinfo {year} {2002})}\BibitemShut {NoStop}%
\bibitem [{\citenamefont {Weides}\ \emph {et~al.}(2006)\citenamefont {Weides},
  \citenamefont {Kemmler}, \citenamefont {Goldobin}, \citenamefont {Koelle},
  \citenamefont {Kleiner}, \citenamefont {Kohlstedt},\ and\ \citenamefont
  {Buzdin}}]{doi:10.1063/1.2356104}%
  \BibitemOpen
  \bibfield  {author} {\bibinfo {author} {\bibfnamefont {M.}~\bibnamefont
  {Weides}}, \bibinfo {author} {\bibfnamefont {M.}~\bibnamefont {Kemmler}},
  \bibinfo {author} {\bibfnamefont {E.}~\bibnamefont {Goldobin}}, \bibinfo
  {author} {\bibfnamefont {D.}~\bibnamefont {Koelle}}, \bibinfo {author}
  {\bibfnamefont {R.}~\bibnamefont {Kleiner}}, \bibinfo {author} {\bibfnamefont
  {H.}~\bibnamefont {Kohlstedt}}, \ and\ \bibinfo {author} {\bibfnamefont
  {A.}~\bibnamefont {Buzdin}},\ }\href {\doibase 10.1063/1.2356104} {\bibfield
  {journal} {\bibinfo  {journal} {Applied Physics Letters}\ }\textbf {\bibinfo
  {volume} {89}},\ \bibinfo {pages} {122511} (\bibinfo {year} {2006})},\
  \Eprint {http://arxiv.org/abs/https://doi.org/10.1063/1.2356104}
  {https://doi.org/10.1063/1.2356104} \BibitemShut {NoStop}%
\bibitem [{\citenamefont {Born}\ \emph {et~al.}(2006)\citenamefont {Born},
  \citenamefont {Siegel}, \citenamefont {Hollmann}, \citenamefont {Braak},
  \citenamefont {Golubov}, \citenamefont {Gusakova},\ and\ \citenamefont
  {Kupriyanov}}]{PhysRevB.74.140501}%
  \BibitemOpen
  \bibfield  {author} {\bibinfo {author} {\bibfnamefont {F.}~\bibnamefont
  {Born}}, \bibinfo {author} {\bibfnamefont {M.}~\bibnamefont {Siegel}},
  \bibinfo {author} {\bibfnamefont {E.~K.}\ \bibnamefont {Hollmann}}, \bibinfo
  {author} {\bibfnamefont {H.}~\bibnamefont {Braak}}, \bibinfo {author}
  {\bibfnamefont {A.~A.}\ \bibnamefont {Golubov}}, \bibinfo {author}
  {\bibfnamefont {D.~Y.}\ \bibnamefont {Gusakova}}, \ and\ \bibinfo {author}
  {\bibfnamefont {M.~Y.}\ \bibnamefont {Kupriyanov}},\ }\href {\doibase
  10.1103/PhysRevB.74.140501} {\bibfield  {journal} {\bibinfo  {journal} {Phys.
  Rev. B}\ }\textbf {\bibinfo {volume} {74}},\ \bibinfo {pages} {140501}
  (\bibinfo {year} {2006})}\BibitemShut {NoStop}%
\bibitem [{\citenamefont {Buzdin}\ \emph {et~al.}(1992)\citenamefont {Buzdin},
  \citenamefont {Bujicic},\ and\ \citenamefont {Kupriyanov}}]{BuzdinJETP1992}%
  \BibitemOpen
  \bibfield  {author} {\bibinfo {author} {\bibfnamefont {A.~I.}\ \bibnamefont
  {Buzdin}}, \bibinfo {author} {\bibfnamefont {B.}~\bibnamefont {Bujicic}}, \
  and\ \bibinfo {author} {\bibfnamefont {M.~Y.}\ \bibnamefont {Kupriyanov}},\
  }\href {http://www.jetp.ac.ru/cgi-bin/e/index/e/74/1/p124?a=list} {\bibfield
  {journal} {\bibinfo  {journal} {Sov. Phys. JETP}\ }\textbf {\bibinfo {volume}
  {74}},\ \bibinfo {pages} {124} (\bibinfo {year} {1992})}\BibitemShut
  {NoStop}%
\bibitem [{\citenamefont {Anh}\ \emph {et~al.}(2016)\citenamefont {Anh},
  \citenamefont {Hai},\ and\ \citenamefont {Tanaka}}]{Anh2016Ncomm}%
  \BibitemOpen
  \bibfield  {author} {\bibinfo {author} {\bibfnamefont {L.~D.}\ \bibnamefont
  {Anh}}, \bibinfo {author} {\bibfnamefont {P.~N.}\ \bibnamefont {Hai}}, \ and\
  \bibinfo {author} {\bibfnamefont {M.}~\bibnamefont {Tanaka}},\ }\href
  {\doibase 10.1038/ncomms13810} {\bibfield  {journal} {\bibinfo  {journal}
  {Nature Communications}\ }\textbf {\bibinfo {volume} {7}},\ \bibinfo {pages}
  {13810} (\bibinfo {year} {2016})}\BibitemShut {NoStop}%
\bibitem [{\citenamefont {Barone}\ \emph {et~al.}(1977)\citenamefont {Barone},
  \citenamefont {Patern\`o}, \citenamefont {Russo},\ and\ \citenamefont
  {Vaglio}}]{doi:10.1002/pssa.2210410206}%
  \BibitemOpen
  \bibfield  {author} {\bibinfo {author} {\bibfnamefont {A.}~\bibnamefont
  {Barone}}, \bibinfo {author} {\bibfnamefont {G.}~\bibnamefont {Patern\`o}},
  \bibinfo {author} {\bibfnamefont {M.}~\bibnamefont {Russo}}, \ and\ \bibinfo
  {author} {\bibfnamefont {R.}~\bibnamefont {Vaglio}},\ }\href {\doibase
  10.1002/pssa.2210410206} {\bibfield  {journal} {\bibinfo  {journal} {physica
  status solidi (a)}\ }\textbf {\bibinfo {volume} {41}},\ \bibinfo {pages}
  {393} (\bibinfo {year} {1977})}\BibitemShut {NoStop}%
\bibitem [{\citenamefont {Barone}\ and\ \citenamefont
  {Paterno}(1982)}]{BaronePaterno198205}%
  \BibitemOpen
  \bibfield  {author} {\bibinfo {author} {\bibfnamefont {A.}~\bibnamefont
  {Barone}}\ and\ \bibinfo {author} {\bibfnamefont {G.}~\bibnamefont
  {Paterno}},\ }\href {http://amazon.co.jp/o/ASIN/0471014699/} {\emph {\bibinfo
  {title} {Physics and Applications of the Josephson Effect}}},\ \bibinfo
  {edition} {1st}\ ed.\ (\bibinfo  {publisher} {Wiley-VCH},\ \bibinfo {year}
  {1982})\BibitemShut {NoStop}%
\bibitem [{\citenamefont {Hikino}\ and\ \citenamefont
  {Yunoki}(2013)}]{PhysRevLett.110.237003}%
  \BibitemOpen
  \bibfield  {author} {\bibinfo {author} {\bibfnamefont {S.}~\bibnamefont
  {Hikino}}\ and\ \bibinfo {author} {\bibfnamefont {S.}~\bibnamefont
  {Yunoki}},\ }\href {\doibase 10.1103/PhysRevLett.110.237003} {\bibfield
  {journal} {\bibinfo  {journal} {Phys. Rev. Lett.}\ }\textbf {\bibinfo
  {volume} {110}},\ \bibinfo {pages} {237003} (\bibinfo {year}
  {2013})}\BibitemShut {NoStop}%
\bibitem [{\citenamefont {Eschrig}\ and\ \citenamefont
  {L\"{o}fwander}(2008)}]{NatPhys.4.138}%
  \BibitemOpen
  \bibfield  {author} {\bibinfo {author} {\bibfnamefont {M.}~\bibnamefont
  {Eschrig}}\ and\ \bibinfo {author} {\bibfnamefont {T.}~\bibnamefont
  {L\"{o}fwander}},\ }\href
  {http://www.nature.com/nphys/journal/v4/n2/abs/nphys831.html} {\bibfield
  {journal} {\bibinfo  {journal} {Nat. Phys.}\ }\textbf {\bibinfo {volume}
  {4}},\ \bibinfo {pages} {138} (\bibinfo {year} {2008})}\BibitemShut {NoStop}%
\end{thebibliography}
\end{document}